\documentstyle[sprocl]{article}

\input{psfig}

\def\Journal#1#2#3#4{{#1} {\bf #2}, #3 (#4)}

\def\ApJ{\em ApJ}
\def\AA{\em A\&A}

\def\be{\begin{equation}}
\def\ee{\end{equation}}
\def\bea{\begin{eqnarray}}
\def\eea{\end{eqnarray}}

\begin{document}


\title{EVOLUTION OF LITHIUM IN THE MILKY WAY}

\author{D. ROMANO}

\address{International School for Advanced Studies, SISSA/ISAS, Via Beirut 
2--4,\\34014 Trieste, Italy\\E-mail: romano@sissa.it} 

\author{F. MATTEUCCI}

\address{Dipartimento di Astronomia, Universit\`a di Trieste, Via G.B. Tiepolo 
11,\\34131 Trieste, Italy\\E-mail: matteucci@ts.astro.it}

\author{P. VENTURA, F. D'ANTONA}

\address{Osservatorio Astronomico di Roma, Via Frascati 33,\\00040 Monte 
Porzio Catone, Italy\\E-mail: paolo, dantona@coma.mporzio.astro.it}


\maketitle\abstracts{ We adopt up-to-date $^7$Li yields from asymptotic giant 
branch (AGB) stars in order to study the temporal evolution of this element 
in the solar neighbourhood. Several lithium stellar sources are considered 
besides the AGBs: Type II supernovae (SNe), novae, low-mass giants. Galactic 
cosmic ray (GCR) nucleosynthesis is taken into account as well. We conclude 
that AGB stars do not substantially contribute to $^7$Li enrichment on a 
Galactic scale. Therefore, a significant $^7$Li production from novae and 
low-mass stars is needed to explain the late, steep rise of the $^7$Li 
abundance in disk stars and the meteoritic $^7$Li abundance.}

\section{Model Prescriptions and Results}

The upper envelope of the observed log\,$\epsilon$($^7$Li) vs. [Fe/H] diagram 
can constrain models of chemical evolution, provided that it actually traces 
the $^7$Li enrichment history of the interstellar medium (ISM) in the solar 
neighbourhood. A nearly flat plateau at low metallicities (the {\em Spite 
Plateau}) is followed by a steep rise afterwards (big dots in Figure 1): these 
features are interpreted as due to $^7$Li enhancement from a lower primordial 
value to a higher present one. The main purpose of this contribution is to 
show how the interplay between several categories of $^7$Li producers may 
explain the rise from the Spite plateau and the meteoritic $^7$Li abundance.

The adopted model assumes that the Galaxy forms out of two main episodes of 
accretion: the first one builds up the halo, the second one the disk. The 
Sun is located at a distance of 8 kpc from the Galactic center, the adopted 
Galactic age is 14 Gyr. Details on the model parameters and basic equations 
can be found in Chiappini {\it et al} \cite{cmg} \cite{cmr}; detailed $^7$Li 
synthesis prescriptions are given in Romano {\it et al} \cite{rmvd}.

In Figure 1, left panel, the contributions to $^7$Li enrichment from different 
stellar sources are shown: {\em i)} AGB stars ({\it short-dashed line}); {\em 
ii)} Type II SNe ({\it dotted line}); {\em iii)} low-mass giants ({\it 
long-dashed line}); and {\em iv)} novae ({\it continuous line}). Model 
predictions in the right panel of Figure 1 are obtained by taking into account 
in the model all these stellar sources plus {\em v)} GCRs.

The main conclusions are that AGBs are not substantial $^7$Li producers on a 
Galactic scale (see also the contributions by P. Ventura and F. D'Antona, 
these proceedings) and that the meteoritic $^7$Li abundance can be reproduced 
only if other stellar sources acting at late times (novae and low-mass stars) 
are contributing a substantial fraction ($\sim$ 60\% in our model) of the 
meteoritic $^7$Li. However, it should be said that the fractional contribution 
of each of these two sources is still very uncertain, mainly due to 
uncertainties in the evolutionary histories of the progenitor stars and in the 
stellar yields.

\begin{figure}[t]
\psfig{figure=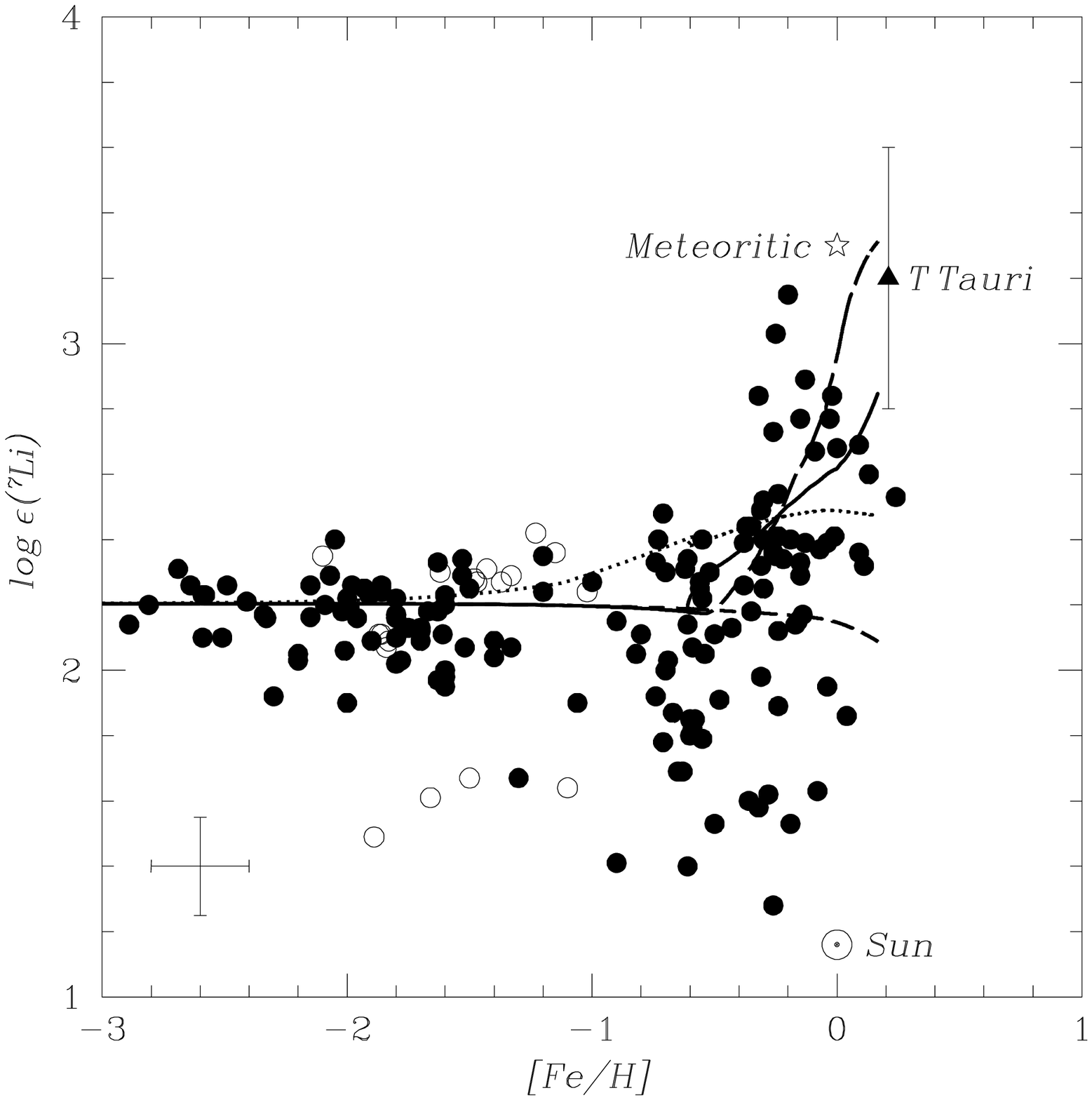,height=5.9cm}
\vskip -5.9cm
\hspace{5.9cm}
\psfig{figure=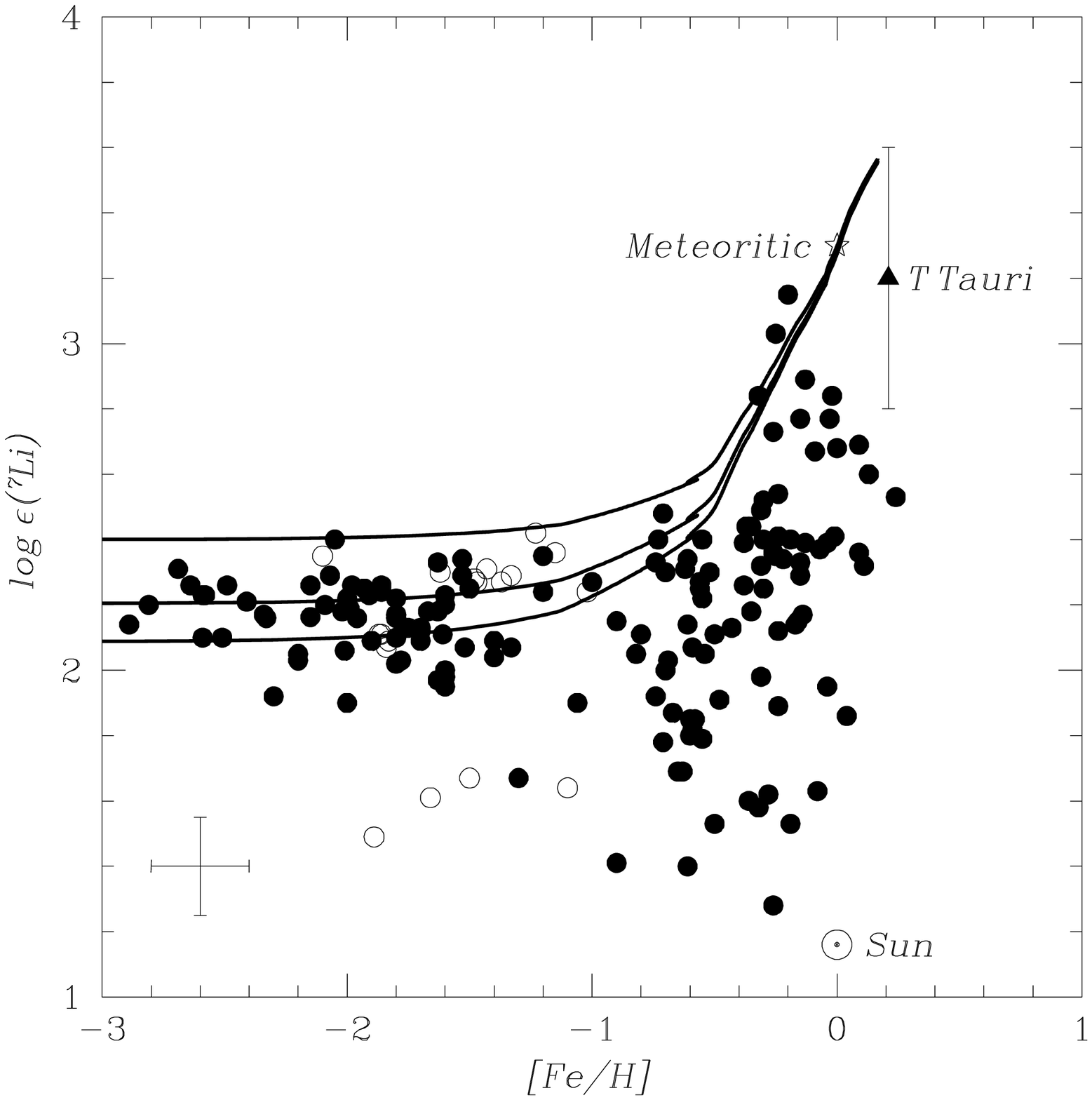,height=5.9cm}
\caption{ Model predictions compared to the data. The contribution from each 
category of stellar $^7$Li producers is shown in the left panel (see text for 
details); the $^7$Li evolution obtained by adding together in the model all 
the stellar sources and GCR nucleosynthesis is shown in the right panel for 
three different values of the primordial abundance. The results at [Fe/H] $>$ 
$-$0.5 dex are nearly the same, owing to the consumption of $^7$Li by stellar 
astration during the Galactic lifetime.}
\end{figure}

\end{document}